# Deep-Learning Tool for Early Identifying Non-Traumatic Intracranial Hemorrhage Etiology based on CT Scan

Meng Zhao[1,2], Yifan Hu, Ruixuan Jiang[2], Yuanli Zhao[1,2], Dong Zhang[1,2], Yan Zhang[1,2], Rong Wang[1,2], Yong Cao[1,2], Qian Zhang[1,2], Yonggang Ma[1,2], Jiaxi Li[1,2], Shaochen Yu[1,2], Wenjie Li[1,2], Ran Zhang[4], Yefeng Zheng[3], Shuo Wang[1,2]*, Jizong Zhao[1,2]*

1. Department of Neurosurgery, Beijing Tiantan Hospital, Capital Medical University, China

2. China National Clinical Research Center for Neurological Diseases, Beijing, China

3. Tencent You Tu Lab, Tencent.

4. Affiliated Hospital of Shandong Jining Medical College

**Correspondence to**

**Jizong Zhao M.D** (Email: zhaojz205@163.com) and **Shuo Wang M.D** (Email: captain9858@vip.sina.com)

# Abstract

**Background**

To develop an artificial intelligence system that can accurately identify acute non-traumatic intracranial hemorrhage (ICH) etiology based on non-contrast CT (NCCT) scans and investigate whether clinicians can benefit from it in a diagnostic setting.

**Materials and Methods**

The deep learning model was developed with 1868 eligible NCCT scans with non-traumatic ICH collected between January 2011 and April 2018. We tested the model on two independent datasets (TT200 and SD 98) collected after April 2018. The model's diagnostic performance was compared with clinicians' performance. We further designed a simulated study to compare the clinicians' performance with and without the deep learning system augmentation.

**Results**

The proposed deep learning system achieved area under the receiver operating curve of 0.986 (95% CI 0.967–1.000) on aneurysms, 0.952 (0.917–0.987) on hypertensive hemorrhage, 0.950 (0.860–1.000) on arteriovenous malformation (AVM), 0.749 (0.586–0.912) on Moyamoya disease (MMD), 0.837 (0.704–0.969) on cavernous malformation (CM), and 0.839 (0.722–0.959) on other causes in TT200 dataset. Given a 90% specificity level, the sensitivities of our model were 97.1% and 90.9% for aneurysm and AVM diagnosis, respectively. The model also shows an impressive generalizability in an independent dataset SD98. The clinicians achieve significant improvements in the sensitivity, specificity, and accuracy of diagnoses of certain hemorrhage etiologies with proposed system augmentation.

## Conclusions

The proposed deep learning algorithms can be an effective tool for early identification of hemorrhage etiologies based on NCCT scans. It may also provide more information for clinicians for triage and further imaging examination selection.

# Introduction

With an annual incidence of 10–30 per 100,000 population, spontaneous and nontraumatic intracerebral hemorrhage (ICH) is the most devastating stroke subtype which accounts for 15%-20% of all strokes and causes significant morbidity and mortality throughout the world.[1],[2] ICH is most commonly caused by hypertension; however, in many cases, ICH is caused by an underlying macrovascular cause, such as arteriovenous malformation (AVM), aneurysm, cavernous malformation (CM), or Moyamoya disease (MMD). [1],[3], [4] Efficient and timely treatments of ICH including early diagnosis and surgical techniques reduce mortality and increase functional survival. [5]  Early and accurate identification of the ICH etiology is vital because therapeutic options vary substantially for different ICH etiologies, especially for underlying vascular abnormalities. [6],[7]

The best strategy for identifying a macrovascular cause in patients with non-traumatic ICH early has not been established.[8] Although DSA is the gold standard for detection of macrovascular abnormalities, it is an invasive procedure associated with risks of complications. The most commonly used emergency room diagnostic tool for patients with symptoms suggesting a hemorrhage is non-contrast computed tomography (NCCT), which offers the advantages of widespread availability, low cost, and rapid acquisition. [6] Following the identification of an ICH by NCCT, computed tomography angiography (CTA) and magnetic resonance angiography (MRA) are recommended to identify underlying vascular lesions. [6] Early diagnosis of underlying vascular abnormalities can affect clinical management and predict prognosis in ICH patients. However, the high cost and limited accessibility of the time-consuming magnetic resonance imaging (MRI), MRA, and CTA generally limit their application to the acute phase. [9] It is rarely possible that every ICH case went through emergency CTA or MRA examination in clinical practice, especially in health care environments for underserved populations. [10] Moreover, there is little guidance on which patients to select for further angiographic imaging for etiology identification. [11] [12]

Although NCCT could be used to screen for macrovascular causes, key features are hard to recognize. Improving the sensitivity, specificity, and accuracy of NCCT screening, with the ability to primarily identify the etiology of ICH, would be crucial in helping physicians to make swift, well-in-formed decisions about selecting for further angiographic imaging or immediately intervention. [13] Convolutional neural networks (CNNs) have shown excellent performance in various clinical image-based recognition tasks and demonstrated potential as a diagnostic strategy. [14] A number of deep learning algorithms based on CNN models have been approved by the US Food and Drug Administration (FDA) for medical image interpretation. [14] Given the potentially less favorable outcomes that from delayed clinical management due to ICH etiology identification process, an accurate and timely deep learning model that could help clinicians reliably identify ICH etiology from NCCT scans is especially valuable. However, to date, the integration of human and artificial intelligence (AI) for ICH etiology detection have barely started, and the ability of deep learning systems to augment clinician performance remains relatively unexplored. [14]

This study aimed to investigate the potential of a CNN system for diagnosis of ICH etiology from NCCT scans. In addition, we also compared the performance of the proposed system with clinician diagnoses. We further conducted a simulated study to compare the clinicians' performance with and without the deep learning system augmentation.

# Methods

## Datasets and clinical taxonomy

**Data collection** For dataset development, we retrospectively reviewed the NCCT scans of 4019 consecutive patients with non-traumatic acute ICH who presented at Beijing Tiantan Hospital between January 2011 and April 2018 (Figure 1). The ICH diagnoses were made independently by two radiologists. Only the first NCCT scan following each ICH event was included in the analysis. We excluded examinations that were: 1) diagnosed as traumatic, 2) associated with a history of brain surgery or external ventricular drain, 3) conducted 24 hours or more after the first ICH event, 4) duplicated scans with rescan due to compromised image quality, initial head motion and metal artifacts arisen from finger and ear rings, 5) lacked a complete image series, or 6) were from patients under four years of age. Medical records were reviewed for patient age, sex, complaint, medical history, known hypertension, and impaired coagulation. Of the original 4019 records, 1868 were eligible for the model development dataset.

To test the accuracy and generalizability of the developed system, we collected two independent test datasets: NCCT scans of consecutive patients with non-traumatic ICH who presented at Beijing Tiantan Hospital from April 2018 to November 2018 (TT200), and NCCT scans of consecutive patients with non-traumatic ICH who presented at the Affiliated Hospital of Shandong Jining Medical College from April 2018 to January 2019 (SD98). Both datasets were subjected to the same exclusion criteria, preprocessing, and labelling methods as the development dataset.

This study was approved by the Institutional Review Boards (IRB) of the Beijing Tiantan Hospital and Affiliated Hospital of Jining Medical College. All medical images and clinical data were fully anonymized. Digital imaging and communications in medicine (DICOM) images were pulled from picture archiving and communication system (PACS) servers in compliance with the Health Insurance Portability and Accountability Act.

**Data labelling** Etiologic classification and definition were based on method proposed by Meretoja et al. [3] Etiology diagnoses were confirmed from NCCT scans in conjunction with further clinical evidence, including CTA, MRA, digital subtraction angiography, or surgical pathology notes. Two radiologists labelled the ICH etiologies, with any disagreement resolved by a third investigator. All radiologists had over 10 years of experience. Each of the NCCT cases were labeled as having one of the six following causes: hypertensive hemorrhage, aneurysm, AVM, MMD, CM, or other causes, which included cerebral amyloid angiopathy, systemic disease, undetermined causes, arteriovenous fistula.

## Data Preprocessing and Model Development

**Data Preprocessing** First, each volume was rotated every 20 degrees along the axial plane with a widely used bilinear algorithm [15], resulting in 18 different image volumes. Second, each volume was resized into a new volume with a consistent physical voxel size of $0.6 \times 0.6 \times 4.2$ $mm^3$, which was almost the same as the mean voxel size of the training set.[16] Third, an unsupervised intensity-based skull-stripping algorithm was applied to the volume, and the stripped volume was cropped to $280 \times 280 \times 30$ voxels. We obtained a total of 33,624 file volumes for the training database.

**Model selection** We compared the classification performance of as AlexNet, [17] ResNet, [18] and SENet, [19] to that of the proposed ICHNet, using the same datasets and processing strategy, and the average classification results are shown in Supplementary Table e-1. AUCs and the overall accuracy served as the evaluation metrics. [20] The proposed model outperformed all of the previous models outperformed on most metrics (See Supplementary Table e-1 and e-3).

**Model development** We used ICHNet for classification of the causes of intracranial hemorrhages from NCCT scans which inspired by CNN architecture SlowFast Networks[21](see Supplementary Figure e-1). The output of ICHNet was a vector for the probability of each cause of ICH. Using stratified random sampling, the initial dataset was split so that both the training and validation sets had similar proportions of the six class labels.

## Training and test protocol

**Training procedure** We oversampled classes other than aneurysm and hypertension because of the imbalanced classes, with the training data for AVM, MMD, CM and others are repeated for 6, 14, 17, and 3 times, respectively. The parameters of the model were initialized by the Kaiming method and optimized with an Adam optimizer. [22] To find the optimal weights, the network was trained by the summation of two different losses, the weighted categorical cross entropy loss and the triplet loss. [23] The weights for the weighted cross entropy loss were computed based on the inverse frequency of each class before oversampling. In the triplet loss function, the triplet margins of these four classes mentioned above were also half of the margins of the two main classes.

**Test procedure** An ensemble strategy was established to validate the proposed approach on the prospective test data by computing the average probability of five models from 5-fold cross validation. The diagnosis could then be predicted from the average probability of all images of the same patient over AUCs. The training and testing procedures were implemented using the Pytorch-0.4.1 package with Python 3.6.

## Performance evaluation

The developed model generated a list of confidence scores ranging from 0 to 1 to indicate the probability of each potential ICH etiology for a given NCCT input. The algorithms were applied to each NCCT case in two independent test datasets. Each data sample was assigned diagnosis based on the largest predicted probability of the model. To evaluate the robustness of the proposed model, overall accuracy was computed and receiver operating characteristic (ROC) curves for each diagnostic label were plotted. Accuracy, sensitivities, specificities, and area under the receiver operating curve (AUCs) were used to assess the algorithms. In addition, one high-sensitivity point and one high-specificity point from the ROC curves for each label were compared with one another (high sensitivity and specificity were both approximately 0.9).

The deep learning model was applied to the two test datasets and the results on TT200 dataset were compared with ICH etiology predictions assigned by six raters (tow radiologists and four neurosurgeons). The two radiologists were neuroradiologists with 7 and 21 years of experience, respectively. Among the four neurosurgeons, two of them had 6 years of experience as attending, one had one year of experience as attending and one had 6 years of experience as resident. None of the raters were involved in data collection or model development.

The raters were instructed to assign one of the six etiology prediction labels to each case. First, three neurosurgeon raters independently assigned ICH etiology predictions based solely on the NCCT images (Task One). After 14 days, the NCCT scans and associated clinical information, including age, sex, chief complaint, physical examination, previous medical and medication history, were given to all six raters to perform the ICH etiology predictions (Task Two). After another 14 days, the raters made etiology predictions again with the knowledge of the proposed model's ICH probabilistic etiology predictions for each case (Task Three). Given the model's confidence score for each etiology, readers had the option to take it into consideration or disregard it based on judgment. Raters were informed of the metrics of the proposed model performance on the validation dataset, but not on the test datasets. All tasks were performed without any time constraint.

## Statistical Analysis

For the model performance assessment, we generated confusion matrices for each of the six etiologies and plotted the true positive rate against the false positive rate for different possible thresholds in one-vs-all diagnostic tests, and the AUC was calculated to evaluate the model. The AUC values for the model were calculated based on the model's prediction scores. The 95% confidence intervals (CIs) of sensitivity and specificity were then calculated using the exact Clopper-Pearson method based on the β distribution,[24] and 95% CIs of AUCs were calculated using the Hanley and McNeil method.[25] The pooled accuracy was obtained from the diagnosis results of all raters. The concordance between paired raters was computed using Cohen κ coefficient. [26] The exact Fleiss κ was also obtained to measure the concordance of all raters.[27] To compare the performances of the algorithm and the raters, we applied the bootstrapping method to obtain samples of metrics for assessment. P-values of less than 0.05 were considered significant.

## Results

For model development, we retrospectively reviewed the NCCT scans consecutive patients with non-traumatic acute ICH who presented at Beijing Tiantan Hospital between January 2011 and April 2018. Of 4019 NCCT scans reviewed, 1868 were eligible for inclusion, as shown in Figure 1. For a qualified input NCCT DICOM volume of an ICH patient, the system was designed as an end-to-end approach, including skull-stripping, rotation and classification, that analyzed the entire NCCT volume automatically and produced a series of scores representing the probabilities of different etiologies (Figure e-2). The Patient demographics and image characteristics from the 1868 records used for model development are summarized in Table 1. In summary, of the 1868 scans of ICH, 628 (33.6%) were diagnosed as related to hypertension, 845 (45.2%) as related to aneurysms, 44 (2.4%) as related to MMD, 34 (1.8%) as related to CMs, 104 (5.6%) as related to AVMs, and 213 (11.4%) as related to other causes. After pre-processing and data augmentation, we had a total of 33,624 NCCT volumes.

The TT200 dataset included 200 examinations. Among these, 75 were labelled as related to hypertension, 70 as aneurysms, 12 as MMD, 11 as AVM, 14 as CM, and 18 as other causes. In the SD98 dataset, 61 records were labeled as related to hypertension, 25 as aneurysms, 5 as MMD, 4 as AVM, 1 as CM, and 2 as other causes (Table 1).

Supplementary Table e-1 and e-2 presented the comparison results of different CNN models for model selection. The performance of the deep learning algorithm ICHNet at the selected operating points are computed, as summarized in Table 2 and Figure 2. On the test dataset TT200, the ICHNet architecture achieved AUC of 0.986 (95% CI 0.967–1.000) on aneurysms, 0.952 (95% CI 0.917–0.987) on hypertensive hemorrhage, 0.950 (95% CI 0.860–1.000) on AVM, 0.749 (95% CI 0.586–0.912) on MMD, 0.837 (95% CI 0.704–0.969) on CMs and 0.839 (95% CI 0.722–0.959) on other causes. On the test dataset SD98, ICHNet achieved AUCs on aneurysms and hypertensive hemorrhage of 0.945 (95% CI 0.882–1.000) and 0.883 (95% CI 0.818–0.948), respectively.

A total of six raters interpreted the TT200 test dataset to evaluate the performance of the proposed deep learning system. Based on the image information only, four of the six clinicians participated in task one and clinicians achieved an average accuracy of 0.706 in task one, in comparison to the proposed approach's accuracy of 0.760. The results of the bootstrapping tests indicated that the proposed system significantly outperformed three of four clinicians (p<0.05), as shown in Supplementary Table e-3 and Figure e-3.

Besides NCCT images, the corresponding clinical information were also provided on task two, and the raters achieved an averaged accuracy of 0.725. Despite the improvement in raters' diagnoses with access to clinical information, the proposed approach still performed significantly better than three of six raters (p<0.05) (Supplementary Figure e-4 and Table e-3).

Given the probability predictions by the algorithms in task three, there were significant increases in the mean sensitivity to aneurysms (0.091, p<0.05) and hypertensive hemorrhage (0.133, p<0.05), and the mean specificity to AVM (0.032, p<0.05), MMD (0.044, p<0.05), and CM (0.028, p<0.05) with AI augmentation on NCCT images and clinical information. Results are presented in Table 3 and Figure 3. Detailed performance of each rater was summarized in Supplementary Table e-4, Table e-5 and Figure e-5. The mean accuracy of each raters' diagnoses was significantly improved from 0.725 to 0.803 (P<0.01) with deep learning augmentation (Figure e-6).

Furthermore, the Cohen's kappa coefficients for pair-wise concordance were evaluated and the results are presented in Figure 4. Before augmentation, the Cohen's kappa coefficients of all pairs of raters were ranged from 0.51 to 0.69. The concordance of all pairs improved to 0.83 (P<0.01) after augmentation. Nine of the 15 pairs of Cohen's kappa coefficients were in an excellent agreement range (>0.75). [28] The Fleiss' kappa for the concordance among all raters was also computed, which increased from 0.61 to 0.75 (P<0.01) after deep learning system augmentation.

# Discussion

To our knowledge, this study is the first to construct a deep learning system that was trained to diagnose ICH etiology from NCCT scans at an accuracy level that rivals human performance. The algorithm was tested on two independent prospective datasets of mixed representative cases. With model augmentation, the sensitivity, specificity, and accuracy of clinicians' etiology diagnoses significantly increased and outperformed the algorithm alone.

Recent research on AI applications that automated medical image diagnosis has demonstrated encouraging outcomes that could lead to fast, cost-effective, and accurate diagnostics that could be accessible worldwide. Deep learning algorithms have been applied across a wide variety of medical scans, including OCT scans [29], retinal fundus images, [30] digitized pathology slides. [31] Chilamkurthy et al. [20] constructed a deep learning algorithm that can accurately identify head CT scan abnormalities requiring urgent attention. Lee et al. [32] developed a deep learning system that detects acute ICH and automatically classifies five ICH locations from NCCT.

Although advances have been made in applying deep learning to ICH CT imaging interpretation, location identification alone is inadequate for clinical practice. Several prediction scores have been described to predict the risk of harboring an underlying macro-vascular etiology in patients with non-traumatic ICH, including the simple ICH score, the secondary ICH score and The DIAGRAM prediction score. [11] However, no such scores could further predict or classify the exact ICH cause. A rapid and accurate diagnosis of the underlying etiology of ICH is essential to direct appropriate management strategies. Our system, which could make further prediction of different etiologies of ICH, could provide more information for clinicians for triage and further imaging selection. [13]

Morever, CT-angiography, which is considered a standard diagnostic tool for ICH etiology, is not routinely performed in most centers. [33] Several studies on spontaneous ICH suggested that about 37.4% - 76.0% patients with ICH underwent further CTA scan. [10, 34, 35] A large study with 1423 consecutive adult patients diagnosed with ICH reported that the sensitivity of CTA for the diagnosis of secondary ICH was 95.7%.[35] For the aneurysm and AVM diagnosis, the sensitivities were 99.1% and 90.4%, respectively. In our study, the algorithm performed well on two independent datasets. Given a 90% specificity level, the sensitivities of our model in TT200 dataset were 97.1% and 90.9% for aneurysm and AVM diagnosis, respectively, which were close to, if not better than, those of CTA.

To compare with the accuracy of raters in a regular clinical setting, clinical information was provided to raters in task two. Although the sensitivity and specificity of the clinicians' diagnoses increased, they were generally still less accurate than the proposed algorithm. Figure e-7 showed illustrative cases that the raters' diagnosis accuracy makes a progress after the foreknowledge of the system's predicted probabilities. When the clinicians were provided with the model's predicted probabilities, they achieved higher specificity in the diagnoses of AVM, MMD, and CM, and sensitivity remained high in the AVM and MMD diagnoses, which was especially important because false negatives of macrovascular etiology diagnoses should be avoided in clinical practice. The sensitivity of raters diagnoses to the hypertensive hemorrhage etiology, the most common ICH etiology, was enhanced sharply with the AI assistance.

This study had some limitations. Our study was hospital-based, and the two datasets were from two large tertiary referral hospitals. The distribution of the hemorrhage etiologies was unbalanced and may have been biased towards referrals. Despite the substantial number of scans with diversified etiologies, the number of cases in minority classes, like MMD and CM, were limited. Therefore, it is important to enrich the training database in the future work, especially the minority classes.

In conclusion, we presented a novel deep learning system that analyzes clinical NCCT scans and makes predictions of ICH etiology with sensitivities, specificities, and accuracies similar to those of clinical specialists. This system could potentially be used as a guide for selecting patients with ICH for neurovascular evaluation to exclude the presence of a vascular abnormality as the ICH etiology. We also demonstrated that integration of the deep learning model can augment clinician performance and could equip specialists with the ability to make better decisions. Further research is necessary to determine the feasibility of applying this algorithm in clinical setting and to determine whether the application could lead to improved care and outcomes compared with current assessment standard.

# Tables

Table 1: Characteristics for both training (TT1868) and two test datasets (TT200 and SD98). AVM = Arteriovenous Malformation; MMD= Moyamoya Disease; CM = Cavernous Malformation CAA = Cerebral Amyloid Angiopathy; AVF = Arteriovenous Fistula; Medication, Anticoagulation-related Medication.

|  | Development dataset | TT200 | SD98 |
|---|---|---|---|
| Number of patients (with both scans and reports) | 1,868 | 200 | 98 |
| Mean age (SD; range) | 52.5±14.9 (4-94) | 50.5±15.9 (4-83) | 55.7±13.0 (15-86) |
| Female patients | 805 (43.1%) | 86 (43.0%) | 37 (37.8%) |
| Etiologies |  |  |  |
|     Aneurysm | 845 (45.2%) | 70 (30.0%) | 25 (25.5%) |
|     Hypertensive hemorrhage | 628 (33.6%) | 75 (32.5%) | 61 (62.2%) |
|     AVM | 104 (5.6%) | 11 (5.5%) | 4 (4.1%) |
|     MMD | 44 (2.4%) | 12 (6.0%) | 5 (5.1%) |
|     CM | 34 (1.8%) | 14 (7.0%) | 1 (1.0%) |
|     Others | 213 (11.4%) | 18 (9.0%) | 2 (2.0%) |
|         Undetermined | 116 (6.2%) | 5 (2.5%) | 0 |
|         Cerebral Venous Thrombosis | 17 (0.9%) | 4 (6.2%) | 0 |
|         CAA | 16 (0.9%) | 4 (6.2%) | 0 |
|         AVF | 9 (0.5%) | 0 | 0 |
|         Medication | 32 (1.7%) | 5 (6.2%) | 2 (2.0%) |
|         Systemic disease/tumor | 23 (1.2%) | 0 | 0 |

Table 2. Performance of algorithms on the TT200 dataset and SD98 dataset. AVM = Arteriovenous Malformation; MMD= Moyamoya Disease; CM = cavernous malformation; AUC=area under the receiver operating characteristic curve

| TT200 dataset | No. of positives | Number of negatives | AUC | High specificity point (=0.9): sensitivity (95% CI) | High sensitivity point (=0.9): specificity (95% CI) |
|---|---|---|---|---|---|
| Aneurysm | 70 | 130 | 0.986 (0.967,1.000) | 0.971 (0.901,0.997) | 0.962 (0.913,0.987) |
| Hypertensive hemorrhage | 75 | 125 | 0.952 (0.917,0.987) | 0.853 (0.753,0.924) | 0.840 (0.764,0.899) |
| AVM | 11 | 189 | 0.950 (0.860,1.000) | 0.909 (0.587,0.998) | 0.910 (0.860,0.947) |
| MMD | 12 | 188 | 0.749 (0.586,0.912) | 0.417 (0.152,0.723) | 0.468 (0.395,0.542) |
| CM | 14 | 186 | 0.837 (0.704,0.969) | 0.571 (0.289,0.823) | 0.457 (0.384,0.532) |
| Others | 18 | 182 | 0.839 (0.722,0.959) | 0.611 (0.358,0.827) | 0.528 (0.452,0.602) |
| SD98 dataset | | | | | |
| Aneurysm | 25 | 73 | 0.945 (0.882,1.000) | 0.904 (0.795,0.952) | 0.920 (0.740,0.990) |
| Hypertensive hemorrhage | 61 | 37 | 0.883 (0.818,0.948) | 0.649 (0.448,0.775) | 0.689 (0.540,0.787) |
| AVM | 4 | 94 | 0.872 | 0.553 | 0.750 |
| MMD | 5 | 93 | 0.796 | 0.172 | 0.800 |
| CM | 1 | 97 | 0.979 | 0.979 | 1.000 |
| Others | 2 | 96 | 0.781 | 0.740 | 0 |

Table 3: Clinician Performance Metrics with and without Augmentation. AVM = Arteriovenous Malformation; MMD= Moyamoya Disease; CM = cavernous malformation; AUC=area under the receiver operating characteristic curve

|  | Sensitivity | | | | Specificity | | | |
| --- | --- | --- | --- | --- | --- | --- | --- | --- |
|  | Without Augmentation | With Augmentation | Increment | p-value | Without Augmentation | With Augmentation | Increment | p-value |
| Aneurysm | 0.874 (0.791,0.957) | 0.964 (0.944,0.984) | 0.091 | 0.019 | 0.949 (0.933,0.964) | 0.947 (0.938,0.957) | -0.001 | 0.744 |
| Hypertensive hemorrhage | 0.798 (0.752,0.843) | 0.931 (0.921,0.942) | 0.133 | 0.002 | 0.917 (0.888,0.946) | 0.896 (0.871,0.921) | -0.021 | 0.926 |
| AVM | 0.697 (0.587,0.807) | 0.712 (0.627,0.797) | 0.015 | 0.500 | 0.920 (0.894,0.946) | 0.952 (0.943,0.960) | 0.032 | 0.018 |
| MMD | 0.264 (0.149,0.379) | 0.236 (0.129,0.343) | -0.028 | 0.690 | 0.943 (0.901,0.986) | 0.988 (0.982,0.993) | 0.044 | 0.008 |
| CM | 0.655 (0.588,0.722) | 0.464 (0.300,0.629) | -0.191 | 0.971 | 0.955 (0.943,0.968) | 0.983 (0.972,0.994) | 0.028 | 0.010 |
| Others | 0.222 (0.121,0.324) | 0.343 (0.219,0.466) | 0.120 | 0.144 | 0.979 (0.961,0.997) | 0.973 (0.96,0.987) | -0.006 | 0.836 |

## Figures

Figure 1. Dataset selection flow chart. (A) Development Dataset. (B) TT200 Dataset. (C) SD98 dataset.

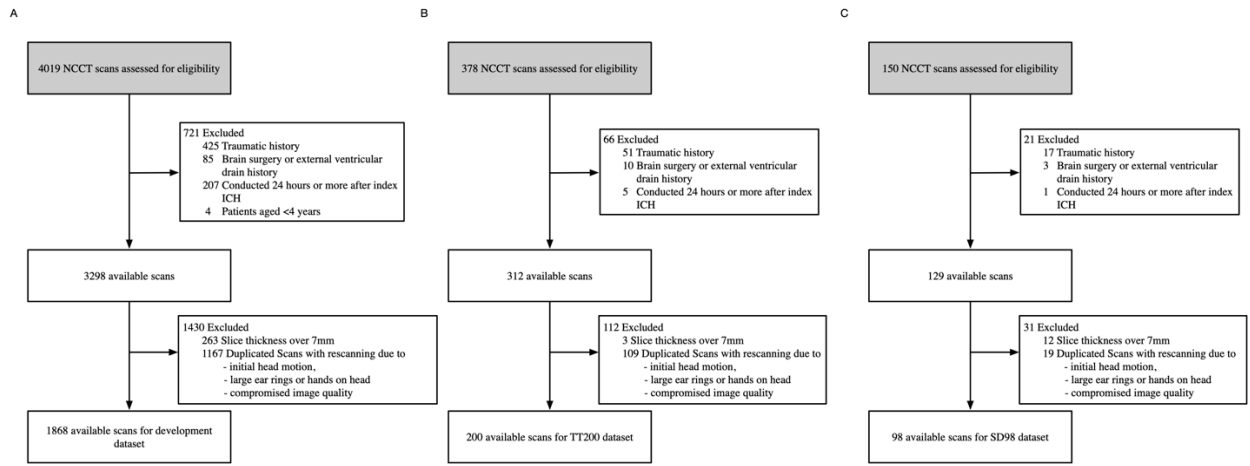

Figure 2: Performance of proposed deep learning system on TT200 and SD98 test datasets.

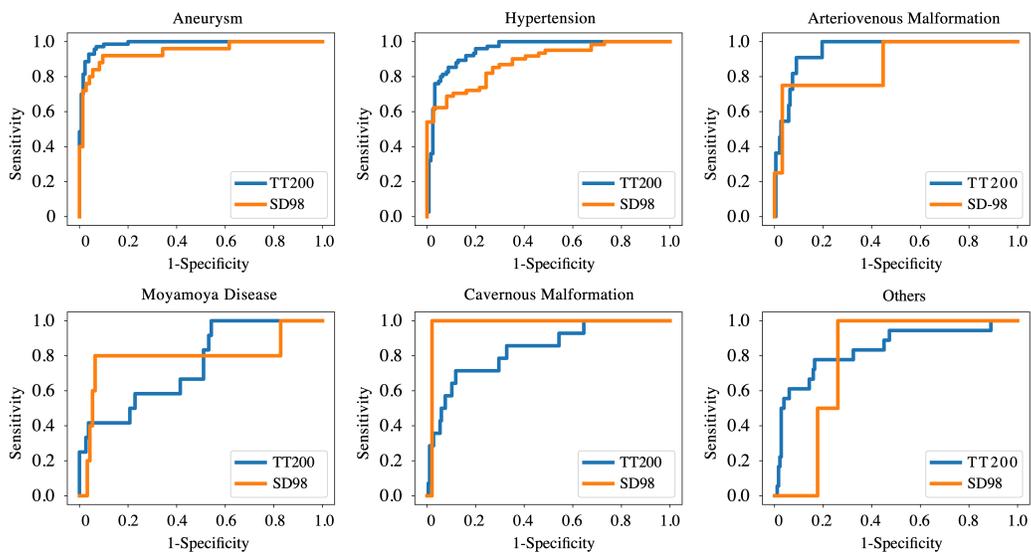

Figure 3: Comparison between ROC curves of ICHNet's and clinicians' sensitivities and specificities before and after augmentation (Task Two and Task Three). ROC, receiver operating characteristic.

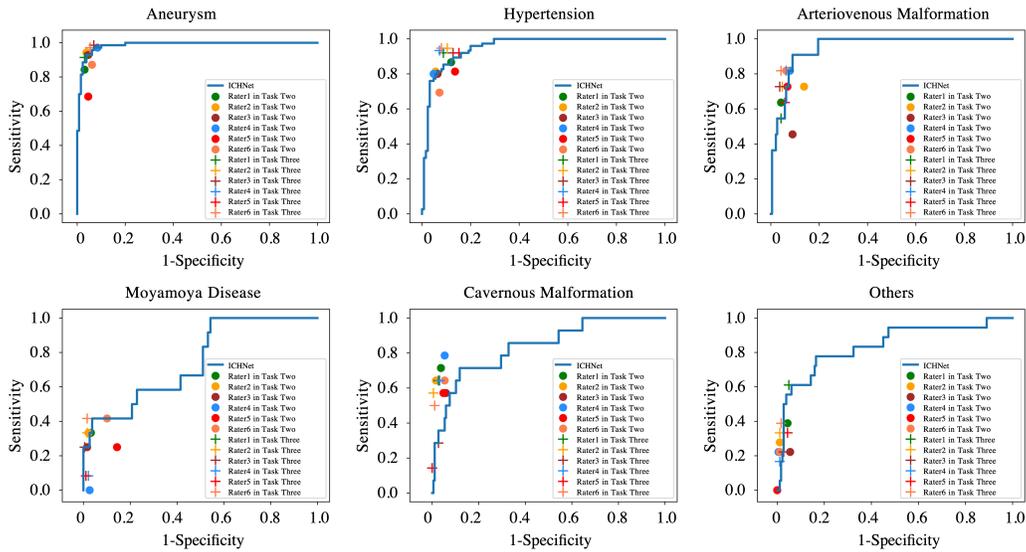

Figure 4: The Cohen's kappa coefficients for pair-wise concordance before (A) and after (B) augmentation.

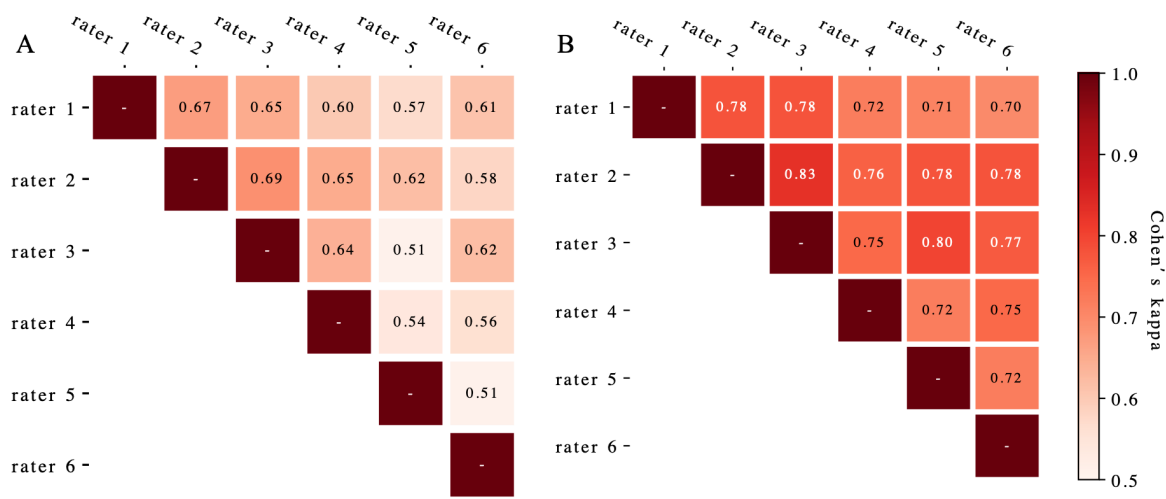

Zhao.    21**Acknowledgments**

This study was funded by The Program of the National Natural Science Foundation of China (81870904). We thank Lu Su, Zhitao Shi, Dan Chen, Haibin Du, and Wenfu Wu.

Zhao. 22# References

1. Qureshi AI, Mendelow AD, Hanley DF (2009) Intracerebral haemorrhage. Lancet 373:1632–1644. doi: 10.1016/S0140-6736(09)60371-8

2. Labovitz DL, Halim A, Boden-Albala B, et al (2005) The incidence of deep and lobar intracerebral hemorrhage in whites, blacks, and Hispanics. Neurology 65:518–522. doi: 10.1212/01.wnl.0000172915.71933.00

3. Meretoja A, Strbian D, Putaala J, et al (2012) SMASH-U: a proposal for etiologic classification of intracerebral hemorrhage. Stroke 43:2592–2597. doi: 10.1161/STROKEAHA.112.661603

4. Lovelock CE, Molyneux AJ, Rothwell PM (2007) Change in incidence and aetiology of intracerebral haemorrhage in Oxfordshire, UK, between 1981 and 2006: a population-based study. Lancet Neurol 6:487–493. doi: 10.1016/S1474-4422(07)70107-2

5. van Donkelaar CE, Bakker NA, Birks J, et al (2020) Impact of Treatment Delay on Outcome in the International Subarachnoid Aneurysm Trial. Stroke 360:STROKEAHA120028993. doi: 10.1161/STROKEAHA.120.028993

6. Hemphill JC, Greenberg SM, Anderson CS, et al (2015) Guidelines for the Management of Spontaneous Intracerebral Hemorrhage: A Guideline for Healthcare Professionals From the American Heart Association/American Stroke Association. Stroke 46:2032–2060. doi: 10.1161/STR.0000000000000069

7. Derdeyn CP, Zipfel GJ, Albuquerque FC, et al (2017) Management of Brain Arteriovenous Malformations: A Scientific Statement for Healthcare Professionals From the American Heart Association/American Stroke Association. Stroke 48:CD003436. doi: 10.1161/STR.0000000000000134

8. Cordonnier C, Demchuk A, Ziai W, Anderson CS (2018) Intracerebral haemorrhage: current approaches to acute management. Lancet 392:1257–1268. doi: 10.1016/S0140-6736(18)31878-6

9. Demchuk AM, Menon BK, Goyal M (2016) Comparing Vessel Imaging. Stroke 47:273–281. doi: 10.1161/STROKEAHA.115.009171

10. Bekelis K, Desai A, Zhao W, et al (2012) Computed tomography angiography: improving diagnostic yield and cost effectiveness in the initial evaluation of spontaneous nonsubarachnoid intracerebral hemorrhage. Journal of Neurosurgery 117:761–766. doi: 10.3171/2012.7.JNS12281

11. Hilkens NA, van Asch CJJ, Werring DJ, et al (2018) Predicting the presence of macrovascular causes in non-traumatic intracerebral haemorrhage: the DIAGRAM prediction score. Journal of Neurology, Neurosurgery & Psychiatry 89:674–679. doi: 10.1136/jnnp-2017-317262

12. Cordonnier C, Klijn CJM, van Beijnum J, Salman RA-S (2010) Radiological Investigation of Spontaneous Intracerebral Hemorrhage. Stroke 41:685–690. doi: 10.1161/STROKEAHA.109.572495

13. Singh MB, Bhatia R (2019) Emergencies in Neurology.

14. Topol EJ (2019) High-performance medicine: the convergence of human and artificial intelligence. Nat Med 1–13. doi: 10.1038/s41591-018-0300-7

15. Ballard DH, Brown CM (1982) Computer vision. englewood cliffs. J: Prentice Hall

16. Han D (2013) Comparison of commonly used image interpolation methods.

17. Krizhevsky A, Sutskever I, Hinton GE (2012) Imagenet classification with deep convolutional neural networks. pp 1097–1105

18. He K, Zhang X, Ren S, Sun J (2015) Deep Residual Learning for Image Recognition. arXiv cs.CV:

19. Hu J, Shen L, Albanie S, et al (2017) Squeeze-and-Excitation Networks. arXiv cs.CV:

20. Chilamkurthy S, Ghosh R, Tanamala S, et al (2018) Deep learning algorithms for detection of critical findings in head CT scans: a retrospective study. Lancet 392:2388–2396. doi: 10.1016/S0140-6736(18)31645-3

21. Feichtenhofer C, Fan H, Malik J, He K (2019) Slowfast networks for video recognition. pp 6202–6211

22. He K, Zhang X, Ren S, Sun J (2016) Deep residual learning for image recognition. pp 770–778

23. Chechik G, Sharma V, Shalit U, Bengio S (2010) Large scale online learning of image similarity through ranking. 11:1109–1135.

24. Clopper CJ, Pearson ES (1934) The use of confidence or fiducial limits illustrated in the case of the binomial. Biometrika 26:404–413.

25. Hanley JA, McNeil BJ (1982) The meaning and use of the area under a receiver operating characteristic (ROC) curve. Radiology 143:29–36. doi: 10.1148/radiology.143.1.7063747

26. Cohen J (2016) A Coefficient of Agreement for Nominal Scales:. Educational and Psychological Measurement 20:37–46. doi: 10.1177/001316446002000104

27. Fleiss JL, Cohen J (1973) The equivalence of weighted kappa and the intraclass correlation coefficient as measures of reliability. Educational and Psychological Measurement 33:613–619.

28. Fleiss JL, Levin B, Paik MC (2013) Statistical methods for rates and proportions. John Wiley & Sons

29. De Fauw J, Ledsam JR, Romera-Paredes B, et al (2018) Clinically applicable deep learning for diagnosis and referral in retinal disease. Nat Med 24:1342–1350. doi: 10.1038/s41591-018-0107-6

30. Gulshan V, Peng L, Coram M, et al (2016) Development and Validation of a Deep Learning Algorithm for Detection of Diabetic Retinopathy in Retinal Fundus Photographs. JAMA 316:2402–9. doi: 10.1001/jama.2016.17216

31. Golden JA (2017) Deep Learning Algorithms for Detection of Lymph Node Metastases From Breast Cancer. JAMA 318:2184–3. doi: 10.1001/jama.2017.14580

32. Lee H, Yune S, Mansouri M, et al (2019) An explainable deep-learning algorithm for the detection of acute intracranial haemorrhage from small datasets. Nature Biomedical Engineering 3:173–182. doi: 10.1038/s41551-018-0324-9

33. Becker KJ, Baxter AB, Bybee HM, et al (1999) Extravasation of radiographic contrast is an independent predictor of death in primary intracerebral hemorrhage. Stroke 30:2025–2032. doi: 10.1161/01.str.30.10.2025

34. Oleinik A, Romero JM, Schwab K, et al (2009) CT Angiography for Intracerebral Hemorrhage Does Not Increase Risk of Acute Nephropathy. Stroke 40:2393–2397. doi: 10.1161/STROKEAHA.108.546127

35. SORIMACHI T, ATSUMI H, YONEMOCHI T, et al (2020) Benefits and Risks of CT Angiography Immediately after Emergency Arrival for Patients with Intracerebral Hematoma. Neurol Med Chir(Tokyo) 60:45–52. doi: 10.2176/nmc.oa.2019-0152

# Supplementary Material

**Table e-1. The accuracy of the different CNN models for model selection.**

|         | An            | Ht            | AVM           | MMD           | CM            | Others        | Pooled Accuracy |
|---------|---------------|---------------|---------------|---------------|---------------|---------------|-----------------|
| AlexNet | 0.9331 (4th)  | 0.9397 (2nd)  | 0.8298 (3rd)  | 0.8113 (2nd)  | 0.8432 (4th)  | 0.7142 (1st)  | 0.7238 (3rd)    |
| ResNet  | 0.9450 (1st)  | 0.9333 (4th)  | 0.8648 (1st)  | 0.7985 (3rd)  | 0.8829 (2nd)  | 0.7127 (2nd)  | 0.7238 (3rd)    |
| SENet   | 0.9360 (3rd)  | 0.9394 (3rd)  | 0.8075 (4th)  | 0.7888 (4th)  | 0.8627 (3rd)  | 0.6969 (4th)  | 0.7383 (1st)    |
| ICHNet  | 0.9395 (2nd)  | 0.9457 (1st)  | 0.8606 (2nd)  | 0.8466 (1st)  | 0.8983 (1st)  | 0.7108 (3rd)  | 0.7383 (1st)    |

**Table e-2. The running time, GPU memory usage, and Parameter Size during experiments for each model.**

|                      | AlexNet       | ResNet      | SENet         | ICHNet      |
|----------------------|---------------|-------------|---------------|-------------|
| Speed (50 iterations)| ~11.5 seconds | ~66 seconds | ~67.5 seconds | ~35 seconds |
| Runtime Memory       | 2087 MB       | 8919 MB     | 9037 MB       | 4441 MB     |
| Parameter Size       | ~58M          | ~133M       | ~133M         | ~135M       |

**Table e-3. Comparison of deep learning and each rater on pooled accuracy in task 1 & 2.**

|               | Task 1           |        | Task 2           |        |
|---------------|------------------|--------|------------------|--------|
|               | Pooled Accuracy  | P      | Pooled Accuracy  | P      |
| Deep Learning | 0.760            |        | 0.760            |        |
| Rater 1       | 0.705            | <0.05  | 0.760            | 0.447  |
| Rater 2       | 0.720            | <0.05  | 0.765            | 0.277  |
| Rater 3       | 0.640            | <0.05  | 0.725            | <0.05  |
| Rater 4       | 0.760            | <0.05  | 0.760            | 0.885  |
| Rater 5       |                  |        | 0.640            | <0.05  |
| Rater 6       |                  |        | 0.700            | <0.05  |

**Table e-4. Overall accuracy performance comparison between unaugmented and augmented results of each rater.**

|             | Rater1 | Rater2 | Rater3 | Rater4 | Rater5 | Rater6 |
|-------------|--------|--------|--------|--------|--------|--------|
| Unaugmented | 0.760  | 0.765  | 0.725  | 0.760  | 0.640  | 0.700  |
| Augmented   | 0.810  | 0.825  | 0.785  | 0.800  | 0.765  | 0.835  |
| Increment   | 0.05   | 0.06   | 0.06   | 0.04   | 0.125  | 0.135  |

**Table e-5. Distribution of CT scanner manufactures in datasets.**

|                     | Siemens | GE  | Phillips |
|---------------------|---------|-----|----------|
| Development dataset | 1512    | 356 | 0        |
| TT200               | 108     | 42  | 50       |
| SD98                | 6       | 92  | 0        |

**Table e-6. Comparison between confusion matrices without and with AI augmentation of each rater.**

| Rater1 ----- Without Augmentation | | | | | | | Rater1 ----- With Augmentation | | | | | |
|---|---|---|---|---|---|---|---|---|---|---|---|---|
| True \ Predict | Aneurysm | Hypertensive hemorrhage | AVM | MMD | CM | Others | Aneurysm | Hypertensive hemorrhage | AVM | MMD | CM | Others |
| Aneurysm | 59 | 2 | 2 | 3 | 0 | 4 | 64 | 2 | 2 | 0 | 0 | 2 |
| Hypertensive hemorrhage | 0 | 65 | 1 | 3 | 5 | 1 | 0 | 69 | 1 | 0 | 4 | 1 |
| AVM | 0 | 2 | 7 | 0 | 0 | 2 | 0 | 2 | 6 | 1 | 0 | 2 |
| MMD | 0 | 5 | 1 | 4 | 1 | 1 | 0 | 4 | 3 | 3 | 0 | 2 |
| CM | 0 | 1 | 3 | 0 | 10 | 0 | 0 | 1 | 2 | 0 | 9 | 2 |
| Others | 4 | 5 | 1 | 0 | 1 | 7 | 4 | 2 | 0 | 0 | 1 | 11 |

| Rater2 ----- Without Augmentation | | | | | | | Rater2 ----- With Augmentation | | | | | |
|---|---|---|---|---|---|---|---|---|---|---|---|---|
| True \ Predict | Aneurysm | Hypertensive hemorrhage | AVM | MMD | CM | Others | Aneurysm | Hypertensive hemorrhage | AVM | MMD | CM | Others |
| Aneurysm | 66 | 0 | 2 | 2 | 0 | 0 | 68 | 0 | 2 | 0 | 0 | 0 |
| Hypertensive hemorrhage | 0 | 61 | 10 | 2 | 2 | 0 | 1 | 71 | 2 | 0 | 1 | 0 |
| AVM | 0 | 1 | 8 | 0 | 1 | 1 | 0 | 2 | 8 | 1 | 0 | 0 |
| MMD | 0 | 4 | 3 | 4 | 0 | 1 | 0 | 6 | 1 | 4 | 0 | 1 |
| CM | 0 | 1 | 4 | 0 | 9 | 0 | 0 | 3 | 2 | 0 | 8 | 1 |
| Others | 5 | 1 | 7 | 0 | 0 | 5 | 6 | 2 | 2 | 2 | 0 | 6 |

| Rater3 ----- Without Augmentation | | | | | | | Rater3 ----- With Augmentation | | | | | |
|---|---|---|---|---|---|---|---|---|---|---|---|---|
| True \ Predict | Aneurysm | Hypertensive hemorrhage | AVM | MMD | CM | Others | Aneurysm | Hypertensive hemorrhage | AVM | MMD | CM | Others |
| Aneurysm | 65 | 0 | 0 | 1 | 2 | 2 | 69 | 0 | 1 | 0 | 0 | 0 |
| Hypertensive hemorrhage | 0 | 60 | 10 | 0 | 4 | 1 | 2 | 69 | 0 | 0 | 4 | 0 |
| AVM | 0 | 0 | 5 | 1 | 2 | 3 | 0 | 2 | 8 | 1 | 0 | 0 |
| MMD | 1 | 4 | 3 | 3 | 1 | 0 | 0 | 5 | 3 | 3 | 0 | 1 |
| CM | 0 | 1 | 0 | 1 | 8 | 4 | 1 | 2 | 3 | 0 | 4 | 4 |
| Others | 5 | 3 | 4 | 0 | 2 | 4 | 6 | 7 | 0 | 0 | 1 | 4 |

| Rater4 ----- Without Augmentation | | | | | | | Rater4 ----- With Augmentation | | | | | |
|---|---|---|---|---|---|---|---|---|---|---|---|---|
| True \ Predict | Aneurysm | Hypertensive hemorrhage | AVM | MMD | CM | Others | Aneurysm | Hypertensive hemorrhage | AVM | MMD | CM | Others |
| Aneurysm | 68 | 0 | 1 | 1 | 0 | 0 | 68 | 1 | 1 | 0 | 0 | 0 |
| Hypertensive hemorrhage | 3 | 60 | 6 | 1 | 5 | 0 | 1 | 70 | 2 | 1 | 1 | 0 |
| AVM | 0 | 0 | 9 | 1 | 1 | 0 | 0 | 1 | 9 | 1 | 0 | 0 |
| MMD | 3 | 2 | 3 | 0 | 4 | 0 | 1 | 4 | 1 | 1 | 5 | 0 |
| CM | 0 | 2 | 0 | 0 | 11 | 1 | 0 | 1 | 2 | 0 | 9 | 2 |
| Others | 5 | 2 | 5 | 2 | 0 | 4 | 5 | 2 | 6 | 2 | 0 | 3 |

| | Rater5 ----- Without Augmentation | | | | | | Rater5 ----- With Augmentation | | | | | |
|---|---|---|---|---|---|---|---|---|---|---|---|---|
| Predict \ True | Aneurysm | Hypertensive hemorrhage | AVM | MMD | CM | Others | Aneurysm | Hypertensive hemorrhage | AVM | MMD | CM | Others |
| Aneurysm | 48 | 2 | 2 | 17 | 1 | 0 | 68 | 1 | 1 | 0 | 0 | 0 |
| Hypertensive hemorrhage | 2 | 61 | 4 | 4 | 4 | 0 | 1 | 69 | 1 | 1 | 0 | 3 |
| AVM | 0 | 1 | 8 | 1 | 1 | 0 | 0 | 3 | 7 | 1 | 0 | 0 |
| MMD | 0 | 6 | 1 | 3 | 2 | 0 | 0 | 6 | 4 | 1 | 0 | 1 |
| CM | 0 | 1 | 4 | 1 | 8 | 0 | 1 | 3 | 4 | 0 | 2 | 4 |
| Others | 4 | 7 | 2 | 4 | 1 | 0 | 5 | 6 | 1 | 0 | 0 | 6 |

| | Rater6 ----- Without Augmentation | | | | | | Rater6 ----- With Augmentation | | | | | |
|---|---|---|---|---|---|---|---|---|---|---|---|---|
| Predict \ True | Aneurysm | Hypertensive hemorrhage | AVM | MMD | CM | Others | Aneurysm | Hypertensive hemorrhage | AVM | MMD | CM | Others |
| Aneurysm | 61 | 1 | 1 | 7 | 0 | 0 | 68 | 0 | 1 | 1 | 0 | 0 |
| Hypertensive hemorrhage | 2 | 52 | 4 | 9 | 8 | 0 | 1 | 71 | 0 | 2 | 1 | 0 |
| AVM | 0 | 1 | 9 | 0 | 0 | 1 | 1 | 1 | 9 | 0 | 0 | 0 |
| MMD | 0 | 3 | 3 | 5 | 1 | 0 | 0 | 5 | 1 | 5 | 1 | 0 |
| CM | 2 | 1 | 0 | 1 | 9 | 1 | 0 | 0 | 4 | 0 | 7 | 3 |
| Others | 4 | 3 | 4 | 2 | 1 | 4 | 5 | 4 | 2 | 0 | 0 | 7 |

**Figure e-1. Schematics of (a) training and (b) test pipelines.**

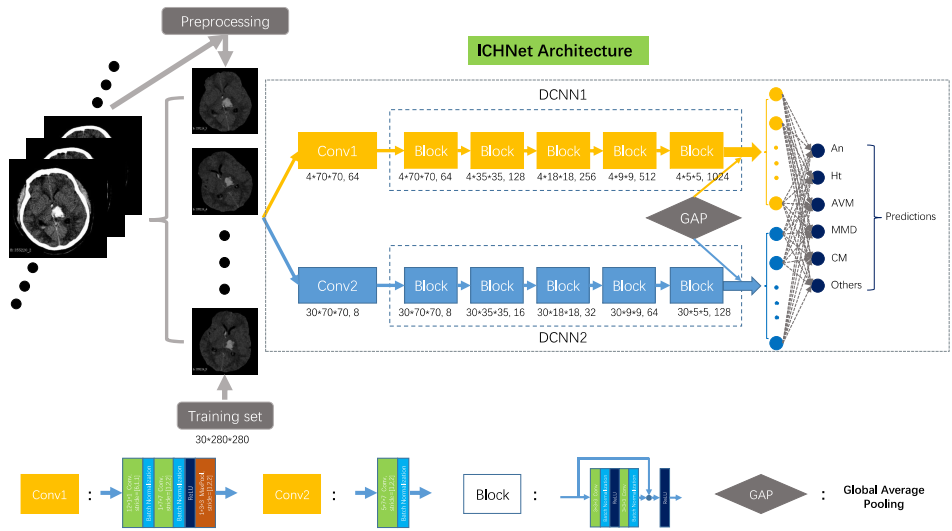

(a)

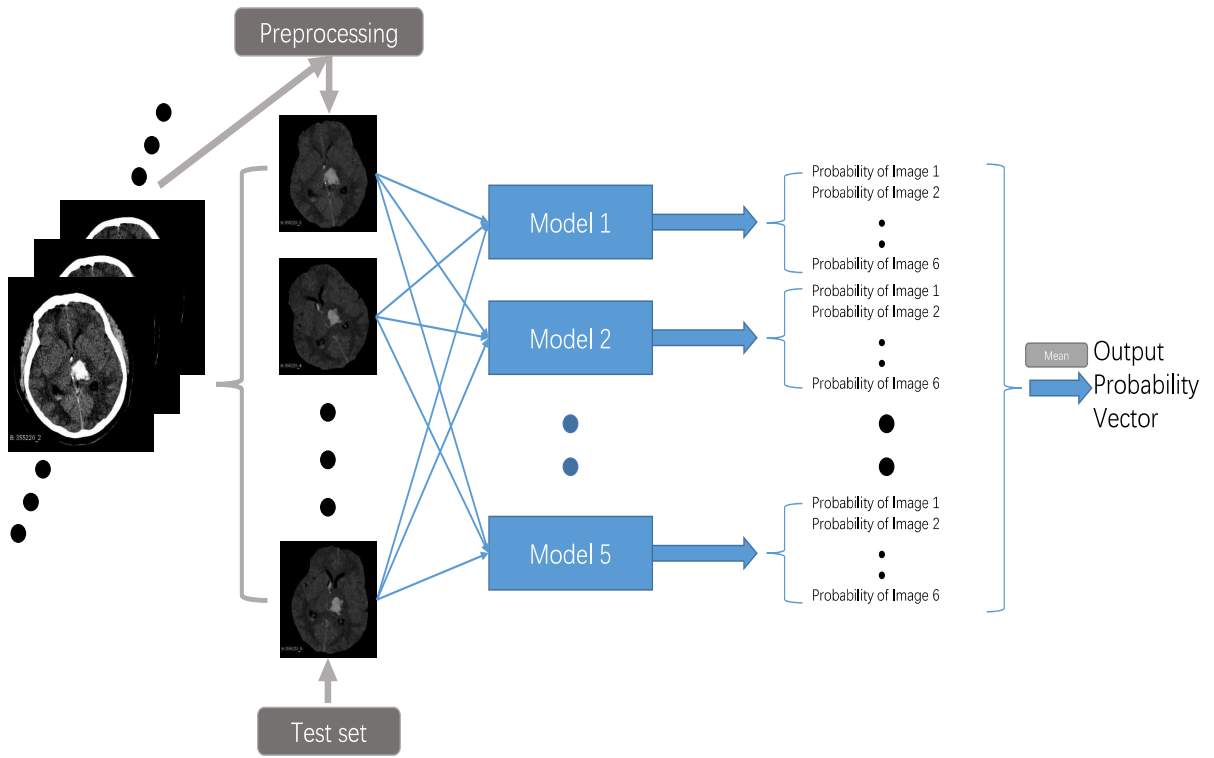

(b)

**Figure e-2. Schematic of data collection and analysis.** (I) Input DICOM images. (II) Randomly divide TT1868 dataset into five folds with four folds in the training set and one-fold in the validation set; TT200 and SD98 were prepared as test sets. (III) Preprocess the DICOM images with skull-stripping and rotation. (IV) Train the models with cross validation strategy to obtain five models. (V) Apply an ensemble strategy to the test set by averaging the outputs of each model. The final output will be the mean score.

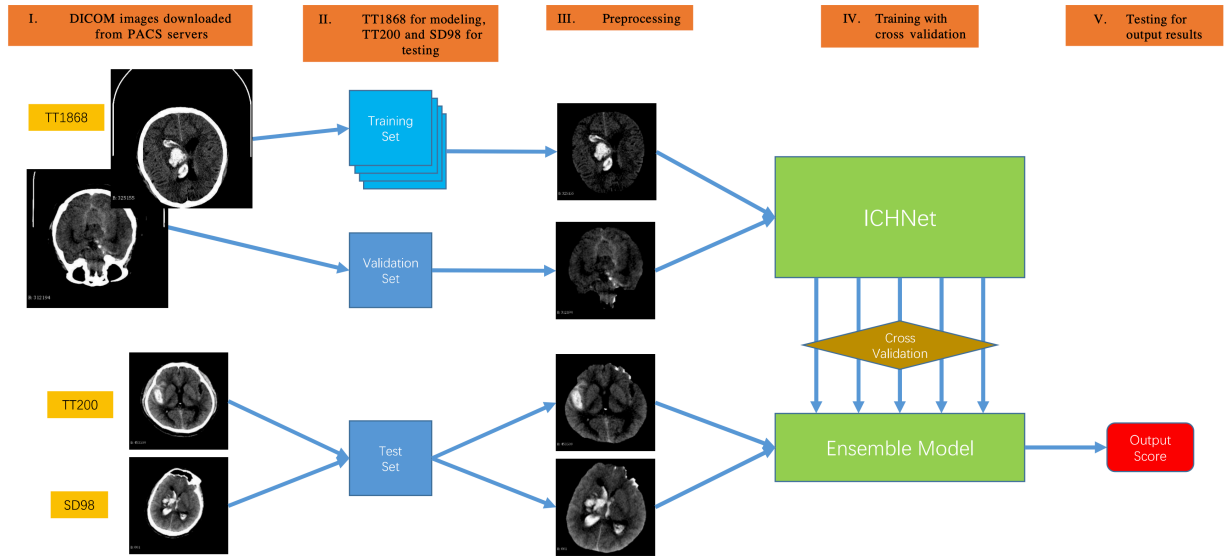

**Figure e-3. ROC curves for the proposed algorithm on the TT200 dataset and the corresponding raters' results from the experiment on this prospective test data.**

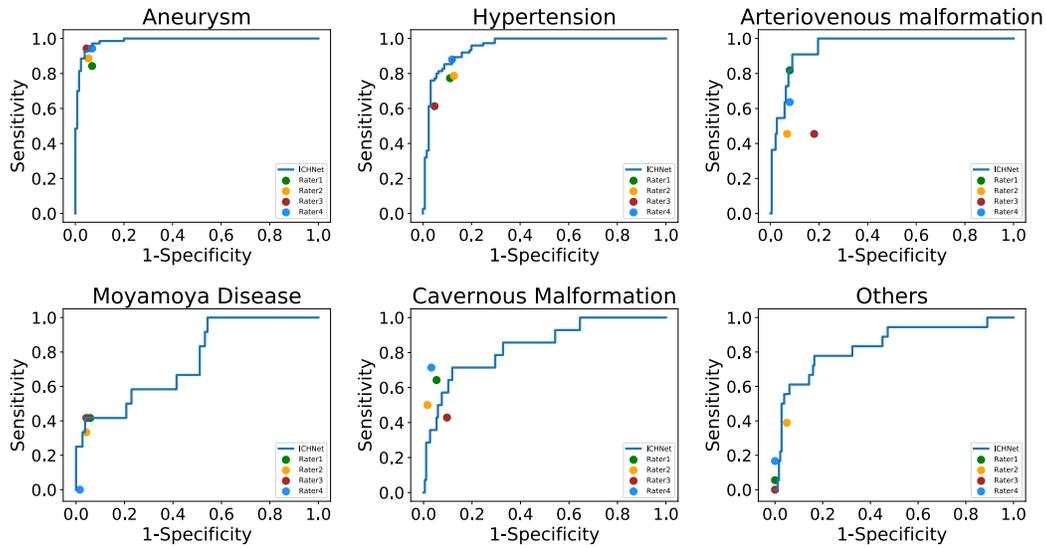

**Figure e-4. ROC curves for the proposed algorithm on the TT200 dataset and the corresponding raters' results from the experiment on this prospective test data, providing the clinical history and NCCT information.**

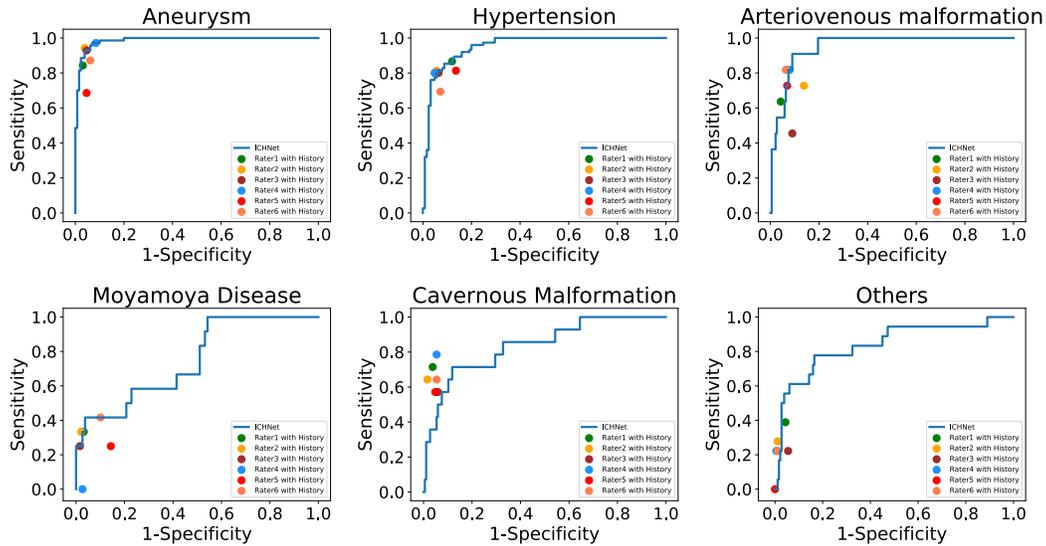

**Figure e-5. Comparison of sensitivities and specificities of each raters' diagnoses without and with AI augmentation.**

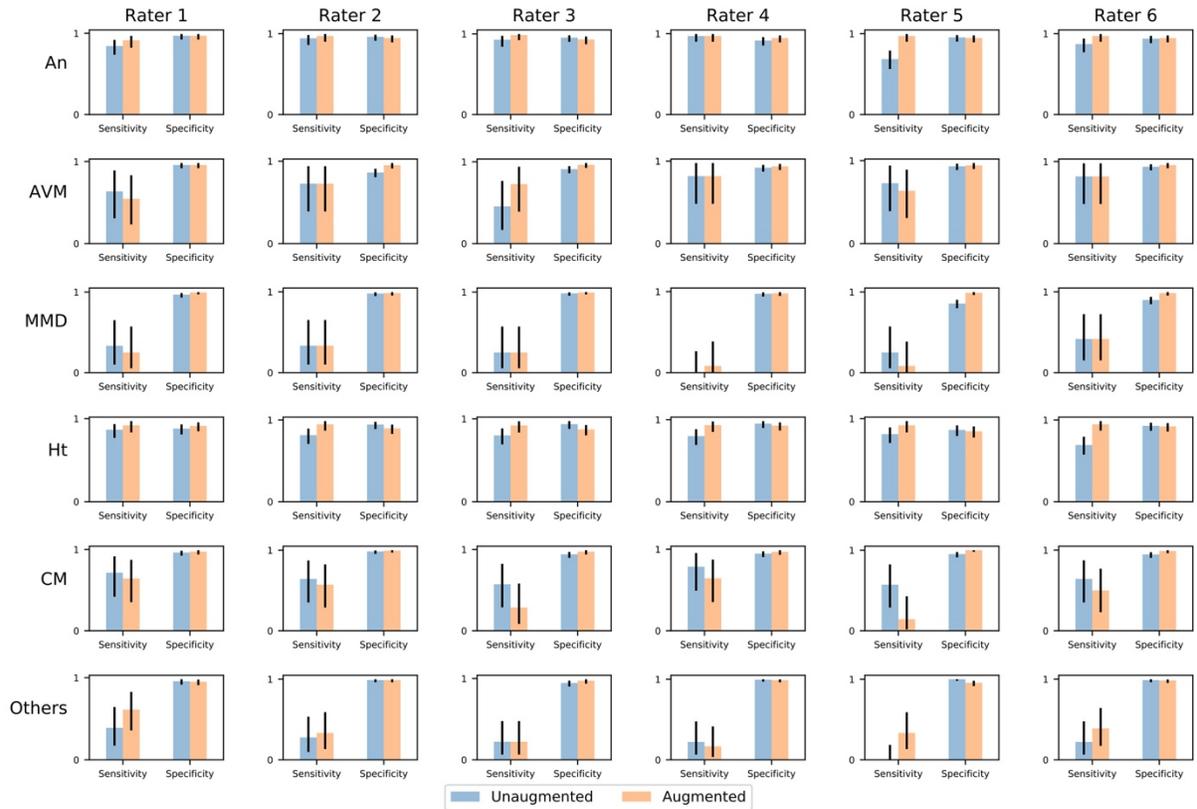

**Figure e-6.** Comparison between accuracy of clinicians' Task one, Task Two, and Task Three diagnoses.

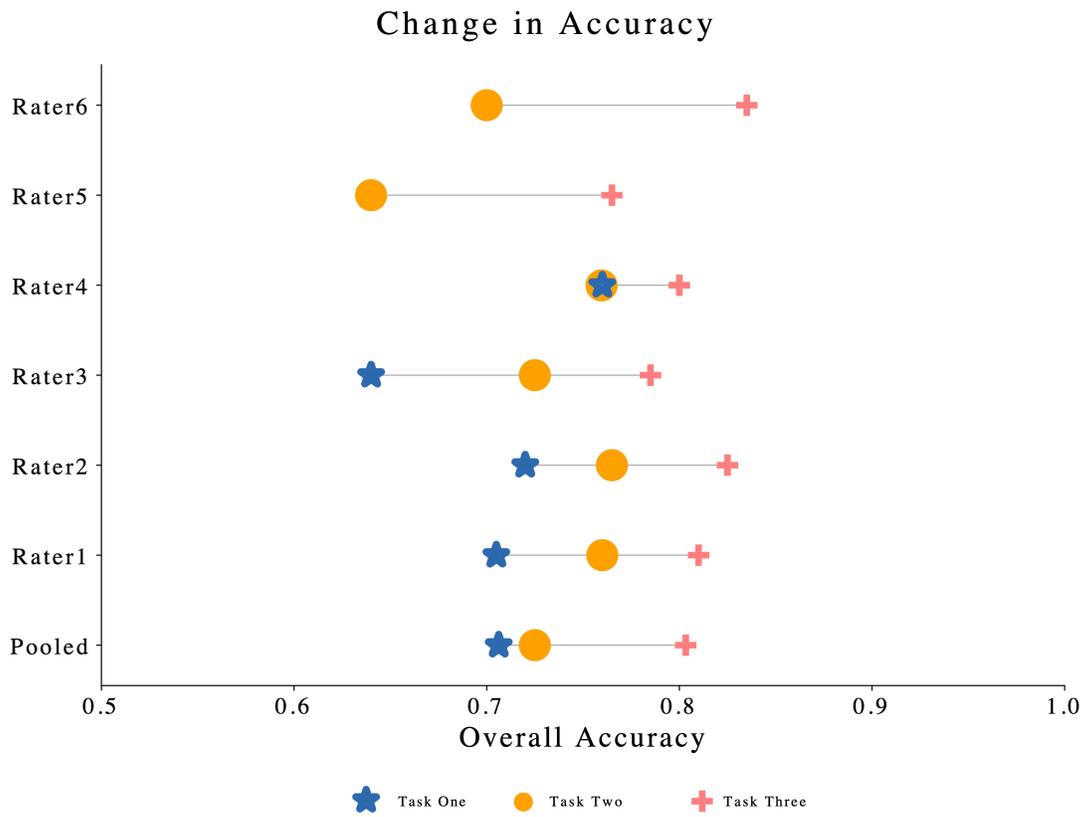

Figure e-7. (a)-(f) show spatial resolution histograms for the training and test datasets. Notably, the distributions of both the intra-slice and inter-slice resolutions of TT200 were closer to the training dataset (TT1868) than to SD98.

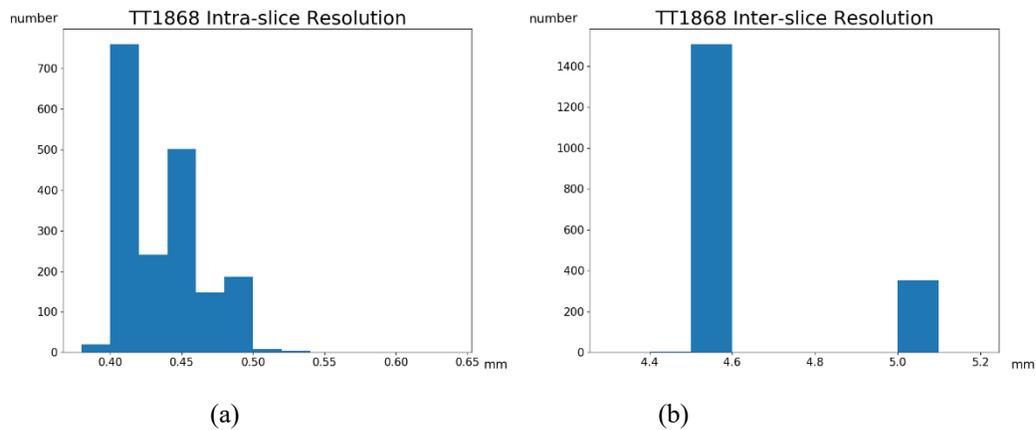

(a)                  (b)

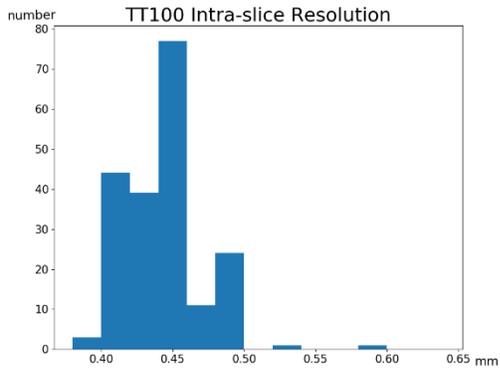
(c)

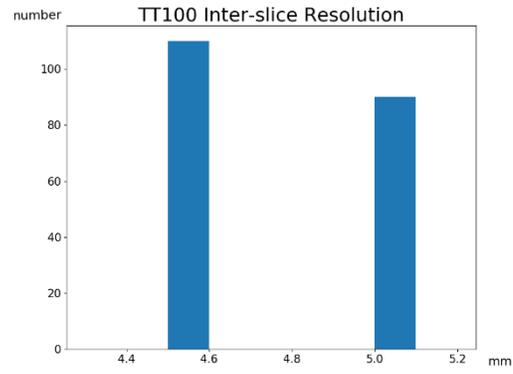
(d)

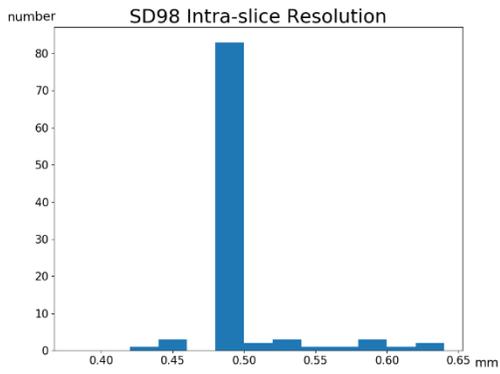
(e)

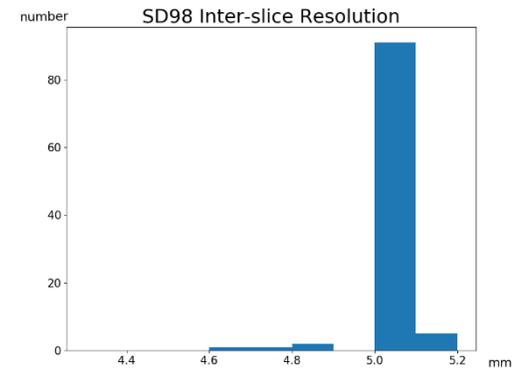
(f)

**Figure e-8. (A)-(D) show images for which the proposed algorithm generated true positive predictions and enhanced the performance of raters with foreknowledge of the predictions. (E) and (F) show images for which the proposed algorithm generated erroneous predictions, but that did not degrade the raters' performance.**

A 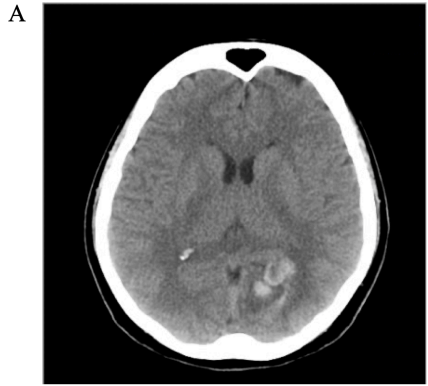

|    | Aneurysm | Hypertensive | AVM | MMD | CM | Others |
|----|----------|--------------|-----|-----|-----|--------|
| AI | 15.4% | 0.8% | 58.5% | 24.4% | 0.6% | 0.3% |
| Ground Truth: AVM | Unaugmented: 3 AVM, 3 CM | | | Augmented: 6 AVM | | |

B 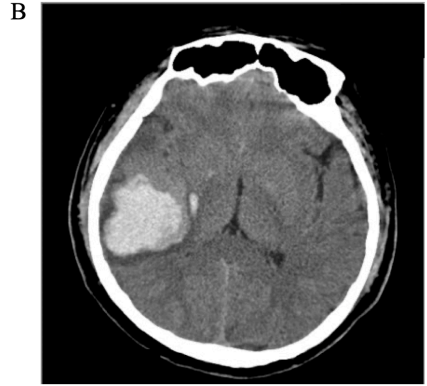

|    | Aneurysm | Hypertensive | AVM | MMD | CM | Others |
|----|----------|--------------|-----|-----|-----|--------|
| AI | <0.1% | 86.7% | 7% | <0.1% | 3.1% | 3.2% |
| Ground Truth: Hypertension | Unaugmented: 6 AVM | | | Augmented: 3 AVM, 3 Hypertension | | |

C 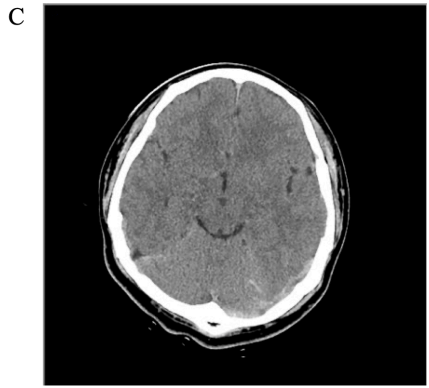

|    | Aneurysm | Hypertensive | AVM | MMD | CM | Others |
|----|----------|--------------|-----|-----|-----|--------|
| AI | 18.2% | 0.2% | 24.1% | <0.1% | 1.1% | 56.4% |
| Ground Truth: Others | Unaugmented: 2 AVM, 1 CM, 1 MMD, 2 Others | | | Augmented: 5 Others, 1 AVM | | |

D 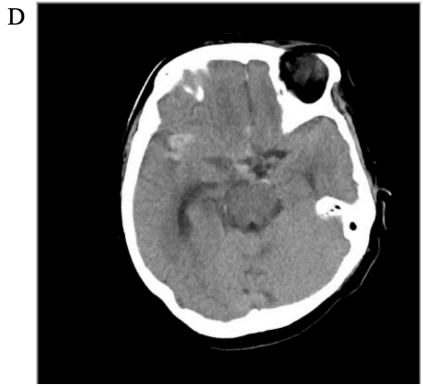

|    | Aneurysm | Hypertensive | AVM | MMD | CM | Others |
|----|----------|--------------|-----|-----|-----|--------|
| AI | 93.5% | 0.1% | <0.1% | <0.1% | 6.3% | <0.1% |
| Ground Truth: Aneurysm | Unaugmented: 3 Aneurysm, 3 MMD | | | Augmented: 6 Aneurysm | | |

E 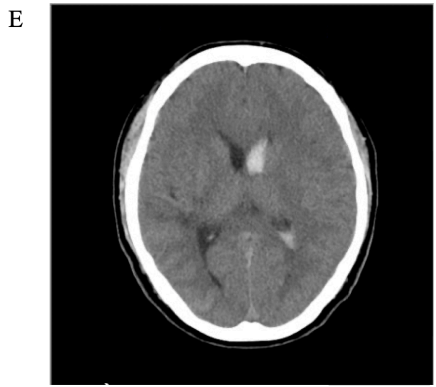

|    | Aneurysm | Hypertensive | AVM | MMD | CM | Others |
|----|----------|--------------|-----|-----|-----|--------|
| AI | 12.2% | 4.9% | 39.6% | 19.5% | 15.2% | 8.7% |
| Ground Truth: MMD | Raters Unaugmented: 4 MMD, 1 Aneurysm, 1 Hypertension | | | Augmented: 4 MMD, 1 AVM, 1 CM | | |

F 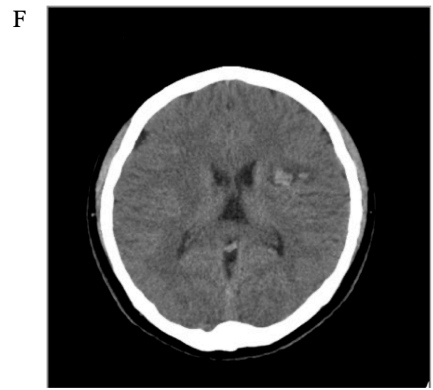

|    | Aneurysm | Hypertensive | AVM | MMD | CM | Others |
|----|----------|--------------|-----|-----|-----|--------|
| AI | 8.9% | <0.1% | 44.3% | 12.7% | 28.4% | 5.7% |
| Ground Truth: CM | Unaugmented: 4 CM, 2 AVM | | | Augmented: 5 CM, 1 AVM | | |